\documentclass[11pt,twoside]{article}
\usepackage{asp2014}

\aspSuppressVolSlug
\resetcounters

\begin{document}

\title{XID+ a new prior-based extraction tool for Herschel-SPIRE maps}

\author{Yannick~Roehlly,$^1$ and Peter D. Hurley$^2$}
\affil{$^1$Aix Marseille Univ, CNRS, LAM, Laboratoire d'Astrophysique de Marseille,
  Marseille, France; \email{yannick.roehlly@lam.fr}}
\affil{$^2$Astronomy Centre, Department of Physics and Astronomy, University of
  Sussex, Falmer, Brighton BN1 9QH, UK}

\paperauthor{Yannick Roehlly}{yannick.roehlly@lam.fr}{0000-0001-8373-8702}{Aix Marseille Univ, CNRS}{LAM, Laboratoire d'Astrophysique de Marseille}{Marseille}{}{13000}{France}
\paperauthor{Peter D. Hurley}{P.D.Hurley@sussex.ac.uk}{}{University of Sussex}{Astronomy Centre, Department of Physics and Astronomy}{Falmer, Brighton}{}{BN1 9QH}{UK}

\begin{abstract}
  We present XID+ a new generation of software for prior-based photometry
  extraction in the Herschel SPIRE maps. Based on a Bayesian framework, XID+
  allows the inclusion of prior information and gives access to the full posterior
  probability distribution of fluxes. XID+ is developed within the Herschel
  Extragalactic Legacy Project (HELP) and is available at
  \url{https://github.com/H-E-L-P/XID_plus}.
\end{abstract}

\section{The Context}

\begin{figure}[htp]
    \centering
    \includegraphics[width=.7\textwidth]{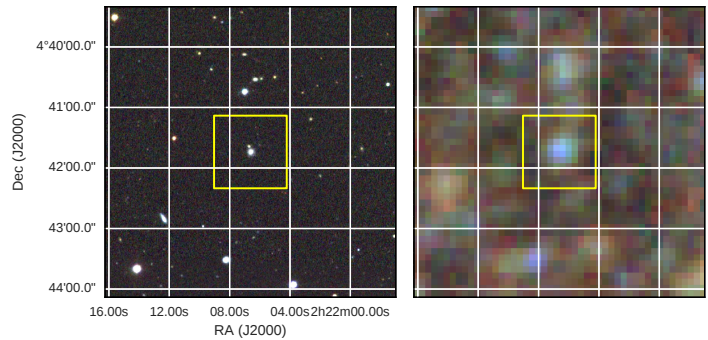}
    \caption{Same part of the sky observed in optical (composite image of SDSS g, r, and
    i bands) and with Herschel (composite image of the three SPIRE bands). The
    square in the centre delimits the zone used in the following figures.
    Because of the large beam, multiple sources cannot be distinguished in the
    SPIRE maps; we refer to this as confusion.}
    \label{fig:P8-21_f1}
  \end{figure}

  ESA's Herschel \emph{Space Laboratory} \citep{2010A&A...518L...1P} has given
  us an unprecedented view of the far-infrared sky. To take most advantage of
  its data, one must nevertheless beat the confusion due to the large beam size
  of it instruments, as illustrated in figure~\ref{fig:P8-21_f1}.

  The \emph{Herschel Extragalactic Legacy Project} (HELP) is a European Research
  Executive Agency funded project that aims to capitalise on the distant
  Universe surveys made by Herschel. To overcome the confusion problem, HELP has
  developed XID+ \citep{2016MNRAS.tmp.1477H}, a new software to perform prior
  based source extraction on confused images.  XID+ is being used on maps from
  Herschel SPIRE and PACS instruments as well as on Spitzer MIPS maps.

\section{Using Bayesian Methods Gives Access to Full Posterior Probability}

\begin{figure}[htp]
    \centering
    \includegraphics[width=.8\textwidth]{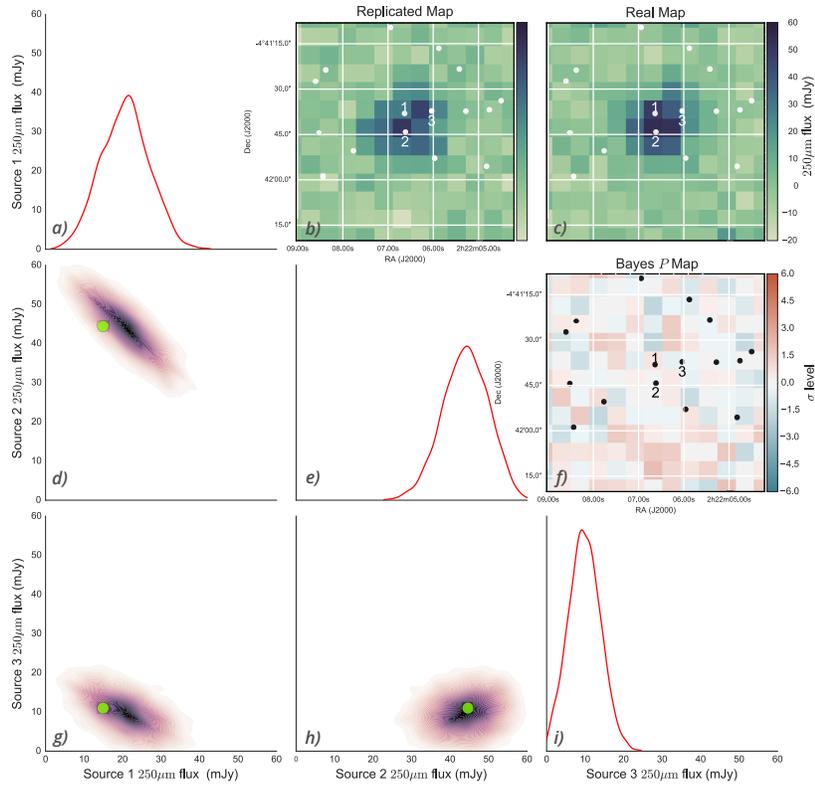}
    \caption{Analysis of the SPIRE $250~\mu{}m$ fluxes of three nearby sources
      with XID+ using only the position as prior information.}
    \label{fig:P8-21_f2}
  \end{figure}

  One way to overcome the confusion is to use information from resolved
  observations, at other wavelengths, that give the positions of known sources.
  XID+ uses Bayesian inference methods implemented within the Stan framework
  \citep{carpenter2016stan} to use this information to compute fluxes.
  Compared to maximum likelihood methods, this gives access to the full
  posterior probability of the flux distribution as illustrated in
  figure~\ref{fig:P8-21_f2}.

  \begin{itemize}
    \item \emph{c)} is the actual SPIRE map;
    \item \emph{d)}, \emph{g)} and \emph{h)} are the joint probability
      distributions of the fluxes for each source pair;
    \item \emph{a)}, \emph{e)} and \emph{i)} are the marginalised probability
      distributions of each source flux;
    \item \emph{b)} is the replicated map corresponding to the green dots on the
      joint distributions.
  \end{itemize}

\section{$p$-Value Maps}

\begin{figure}[htp]
    \centering
    \includegraphics[width=.8\textwidth]{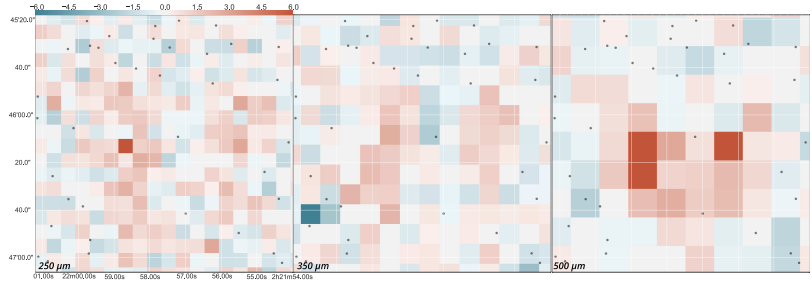}
    \caption{\emph{p-value} maps of the same area for SPIRE~$250$, $350$, and $500~\mu{}m$.}
    \label{fig:P8-21_f3}
  \end{figure}

  One interesting output of XID+ is the \emph{p-value} map. It indicates how
  well the real map is explained by the model and shows unexpected excesses or
  lacks in fluxes. Figure~\ref{fig:P8-21_f3} shows a zone with an unexplained
  excess in the SPIRE~$500~\mu{}m$ filter that may reveal some interesting
  objects not present in the original catalogues.

\section{Adding More Prior Information}

\begin{figure}[htp]
    \centering
    \includegraphics[width=\textwidth]{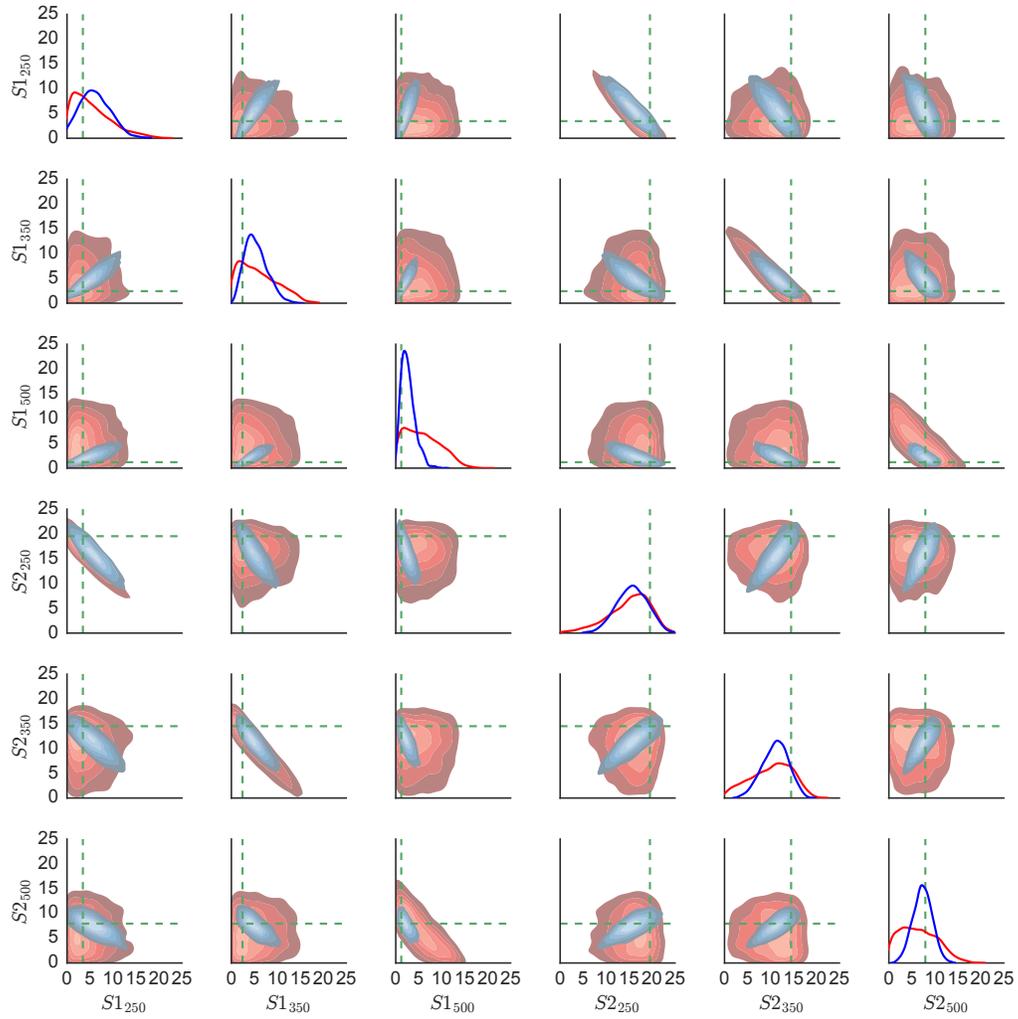}
    \caption{Effect of adding redshift value and simple SEDs as prior information.}
    \label{fig:P8-21_f4}
  \end{figure}

  The use of a Bayesian framework makes it possible to add new prior
  information. For instance, we can use our prior knowledge on redshifts,
  combined with some simple spectral energy distributions (SEDs) to better
  constrain the fluxes by eliminating impossible combinations. This is
  illustrated by figure~\ref{fig:P8-21_f4}: the red (lighter grey in
  black and white) probability density functions (PDFs) are those that don't use
  the redshift and SEDs as prior, the blue (darker grey) are those that use it.

  \acknowledgements The research leading to these results has received funding
  from the European Union Seventh Framework Programme FP7/2007--2013/ under
  grant agreement 607254. \textit{This publication reflects only the author's
  view and the European Union is not responsible for any use that may be made
  of the information contained therein.}\\
  This research made use of Astropy, a community-developed core Python package
  for Astronomy \citep{2013A&A...558A..33A}. This research made use of APLpy,
  an open-source plotting package for Python hosted at
  \url{http://aplpy.github.com}.

\bibliography{P8-21}  

\begin{thebibliography}{}
\expandafter\ifx\csname natexlab\endcsname\relax\def\natexlab#1{#1}\fi
\expandafter\ifx\csname url\endcsname\relax
  \def\url#1{\texttt{#1}}\fi
\expandafter\ifx\csname urlprefix\endcsname\relax\def\urlprefix{URL }\fi
\providecommand{\eprint}[2][]{\url{#2}}

\bibitem[{{Astropy Collaboration}(2013)}]{2013A&A...558A..33A}
{Astropy Collaboration} 2013, \aap, 558, A33. \eprint{1307.6212}

\bibitem[{Carpenter et~al.(2016)Carpenter, Gelman, Hoffman, Lee, Goodrich,
  Betancourt, Brubaker, Guo, Li, \& Riddell}]{carpenter2016stan}
Carpenter, B., Gelman, A., Hoffman, M., Lee, D., Goodrich, B., Betancourt, M.,
  Brubaker, M.~A., Guo, J., Li, P., \& Riddell, A. 2016, J Stat Softw

\bibitem[{{Hurley} et~al.(2016){Hurley}, {Oliver}, {Betancourt}, {Clarke},
  {Cowley}, {Duivenvoorden}, {Farrah}, {Griffin}, {Lacey}, {Le Floc'h},
  {Papadopoulos}, {Sargent}, {Scudder}, {Vaccari}, {Valtchanov}, \&
  {Wang}}]{2016MNRAS.tmp.1477H}
{Hurley}, P.~D., {Oliver}, S., {Betancourt}, M., {Clarke}, C., {Cowley}, W.~I.,
  {Duivenvoorden}, S., {Farrah}, D., {Griffin}, M., {Lacey}, C., {Le Floc'h},
  E., {Papadopoulos}, A., {Sargent}, M., {Scudder}, J.~M., {Vaccari}, M.,
  {Valtchanov}, I., \& {Wang}, L. 2016, \mnras. \eprint{1606.05770}

\bibitem[{{Pilbratt} et~al.(2010){Pilbratt}, {Riedinger}, {Passvogel}, {Crone},
  {Doyle}, {Gageur}, {Heras}, {Jewell}, {Metcalfe}, {Ott}, \&
  {Schmidt}}]{2010A&A...518L...1P}
{Pilbratt}, G.~L., {Riedinger}, J.~R., {Passvogel}, T., {Crone}, G., {Doyle},
  D., {Gageur}, U., {Heras}, A.~M., {Jewell}, C., {Metcalfe}, L., {Ott}, S., \&
  {Schmidt}, M. 2010, \aap, 518, L1. \eprint{1005.5331}

\end{thebibliography}

\end{document}